\documentclass{pasj00}

\def\farcs{\hbox{$.\!\!^{\prime\prime}$}}
\def\arcsec{\hbox{$^{\prime\prime}$\,}}
\newcommand\ion[2]{#1$\;${\scshape{#2}}}%

\begin{document}
\SetRunningHead{Orozco Su\'arez et al.}{Strategy for the inversion of Hinode spectropolarimetric 
measurements in the quiet Sun}
\Received{2007/07/15} 
\Accepted{2007/09/13}

\title{Strategy for the inversion of Hinode spectropolarimetric measurements in the quiet Sun}

\author{David \textsc{Orozco Su\'arez}\altaffilmark{1},
    Luis R.\ \textsc{Bellot Rubio}\altaffilmark{1}, 
    Jose Carlos \textsc{del Toro Iniesta}\altaffilmark{1},
    Saku \textsc{Tsuneta}\altaffilmark{2},
    Bruce W.\ \textsc{Lites}\altaffilmark{3},
    Kiyoshi \textsc{Ichimoto}\altaffilmark{2},
    Yukio \textsc{Katsukawa}\altaffilmark{2},
    Shin'ichi \textsc{Nagata}\altaffilmark{4},
    Toshifumi \textsc{Shimizu}\altaffilmark{5},
    Richard A.\ \textsc{Shine}\altaffilmark{6},
    Yoshinori \textsc{Suematsu}\altaffilmark{2},
    Theodore D.\ \textsc{Tarbell}\altaffilmark{6},
    and
    Alan M.\ \textsc{Title}\altaffilmark{6}}

\altaffiltext{1}{Instituto de Astrof\'{\i}sica de Andaluc\'{\i}a
(CSIC), Apdo.\ de Correos 3004, 18080 Granada, Spain; orozco@iaa.es}

\altaffiltext{2}{National Astronomical Observatory of Japan, 2-21-1 Osawa,
Mitaka, Tokyo 181-8588, Japan}

\altaffiltext{3}{High Altitude Observatory, National Center
for Atmospheric Research, P.O. Box 3000, Boulder, CO 80307, USA}

\altaffiltext{4}{Hida Observatory, Kyoto University, Kamitakara, Gifu
506-1314, Japan} 

\altaffiltext{5}{Institute of Space and Astronautical Science, Japan Aerospace
Exploration Agency, Tokyo, Japan} 

\altaffiltext{6}{Lockheed Martin Solar and Astrophysics Laboratory, B/252,
3251 Hanover St., Palo Alto, CA 94304, USA}

\maketitle

%

\begin{abstract}
In this paper we propose an inversion strategy for the analysis of
spectropolarimetric measurements taken by {\em Hinode} in the quiet Sun.  The
spectropolarimeter of the Solar Optical Telescope aboard {\em Hinode} records
the Stokes spectra of the \ion{Fe}{i} line pair at 630.2~nm with
unprecendented angular resolution, high spectral resolution, and high
sensitivity. We discuss the need to consider a {\em local} stray-light
contamination to account for the effects of telescope diffraction. The
strategy is applied to observations of a wide quiet Sun area at disk
center. Using these data we examine the influence of noise and initial guess
models in the inversion results. Our analysis yields the
distributions of magnetic field strengths and stray-light factors. They show
that quiet Sun internetwork regions consist mainly of hG fields with
stray-light contaminations of about 0.8.
\end{abstract}

%

  \section{Introduction}
  \label{sec:intro}

{\em Hinode} (Kosugi et al.\ 2007) is the first solar satellite carrying a
full vector spectropolarimeter (SP; Lites et al.\ 2001). Since its launch in
September 2006, the instrument has been taking high-precision, high-angular
resolution measurements of the \ion{Fe}{i} lines at 630.2~nm. With a pixel
size of 0\farcs16, the angular resolution achieved by the SP is about
0\farcs32, very close to the diffraction limit of the {\em Hinode} 0.5m Solar
Optical Telescope (Suematsu et al.\ 2007; Tarbell et al.\ 2007; Tsuneta et
al.\ 2007). Such an unprecedented spatial resolution opens exciting
possibilities for the analysis of the weak magnetic signals observed in the
quiet Sun. It should permit, for instance, a better isolation of the magnetic
elements that form the quiet photosphere, provided they are not
organized on scales much smaller than 0\farcs1. The increased spatial
resolution may result in significantly larger polarization signals than
those recorded on the ground. This would minimize the influence of noise,
which has long been recognized as one of the main problems in the study of
quiet Sun magnetic fields.

The availability of very high angular resolution observations,
virtually free from seeing effects, is also important for other
reasons. Since the light entering one pixel comes from a much smaller
region of the solar surface, the effect of different atmospheres
contributing to the intensity and polarization profiles is
decreased. This should facilitate the interpretation of the
measurements, as relatively simple one-component atmospheres may be
sufficient to explain the observations. Stokes inversions of
ground-based data are usually performed in terms of two-component
atmospheres because the intensity and polarization profiles are not
compatible with the signals emerging from a homogeneous magnetic
atmosphere, due to the relatively modest angular resolution attained.

Both the smaller influence of noise and the possibility of using
simple model atmospheres make high resolution measurements ideal to
study the magnetism of the quiet solar photosphere. Space-borne
observations may also shed light on the capabilities of the visible
\ion{Fe}{i} lines at 630.2~nm, which have recently been put into
question by Mart\'{\i}nez Gonz\'alez et al.\ (2006a,b). First attempts
to clarify the diagnostic potential of these lines at high spatial
resolution have been carried out by Khomenko et al.\ (2007a,b). In a
previous paper (Orozco Su\'arez et al.\ 2007), we have investigated
the same problem using radiative magnetoconvection calculations. We
simulated {\em Hinode}/SP measurements and inverted them in terms of
one-component atmospheres accounting for stray/scattered light.  The
main result of those tests is that Milne-Eddington (ME) inversions of
high-angular resolution \ion{Fe}{i} 630.2~nm measurements do recover
the actual field strengths present in the simulation snapshots. We
therefore suggested that the spectropolarimeter aboard {\em Hinode}
would be appropriate for quiet Sun magnetism studies.

\begin{figure*}
\begin{center} 
\FigureFile(6cm,4cm){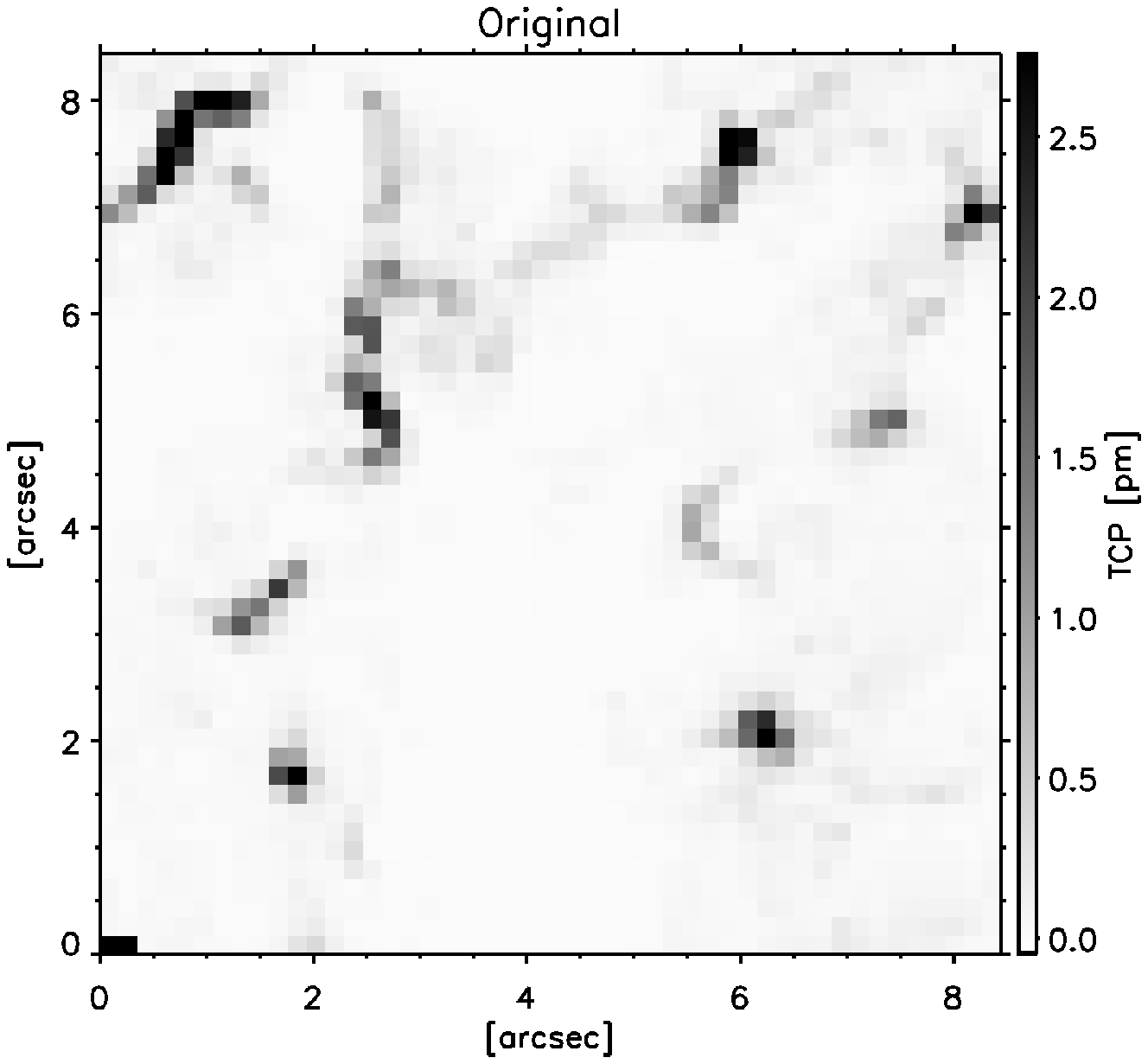} 
\FigureFile(6cm,4cm){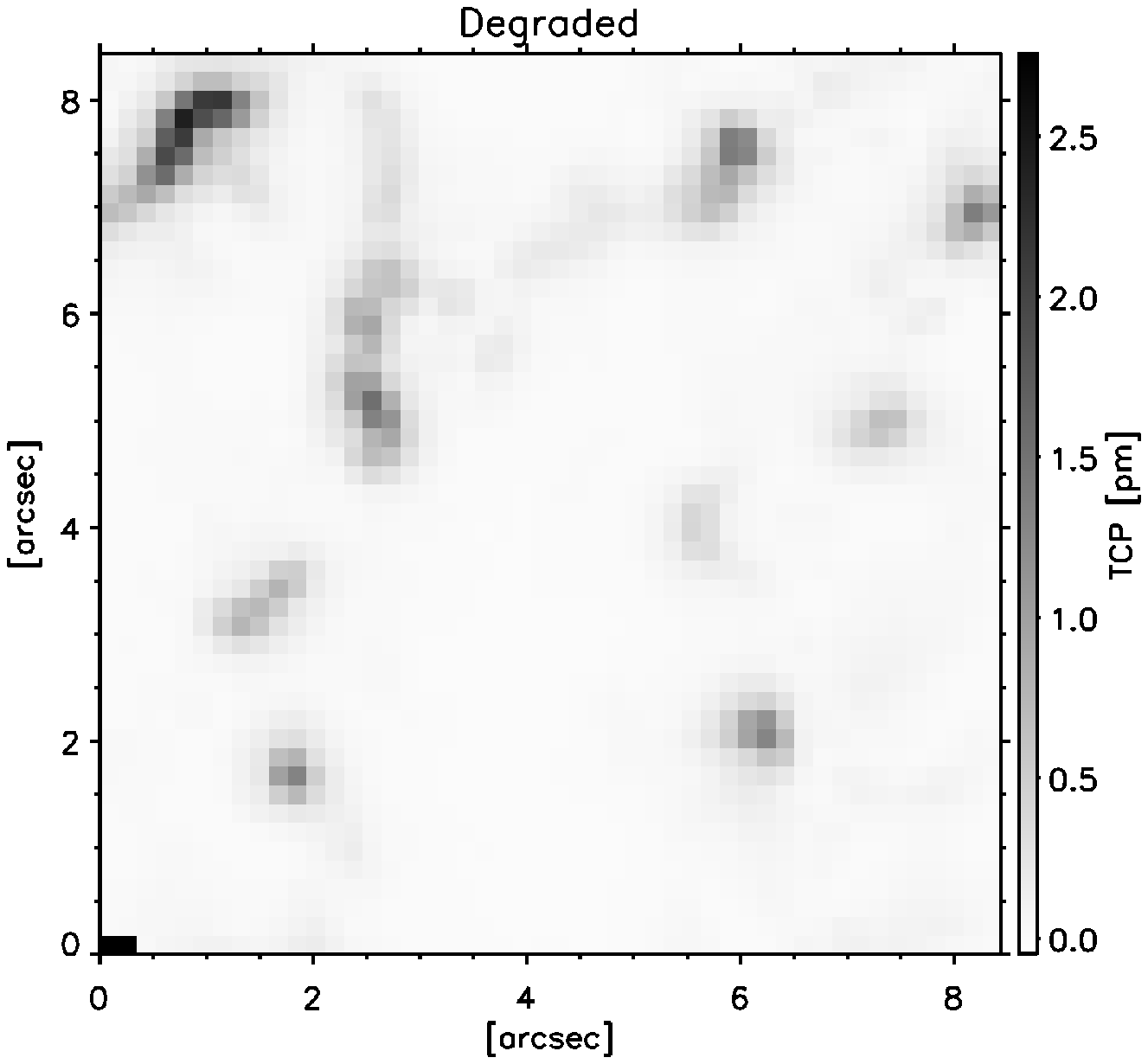} 
\FigureFile(5cm,4cm){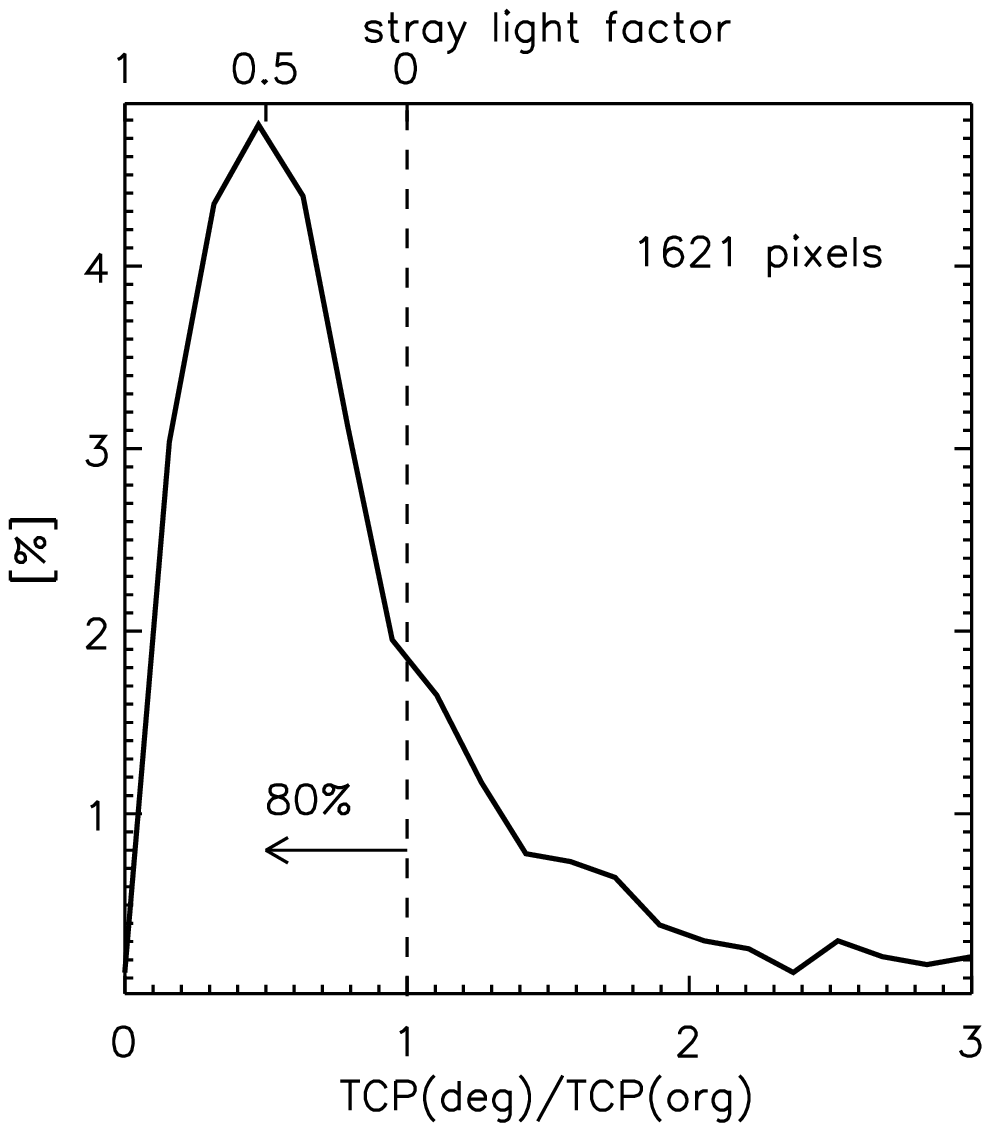} 
\end{center}
\caption{Total circular polarization signals, ${\rm TCP} = \int |V| \, {\rm
d}\lambda/I_{\rm c}$, in the original ({\em left}) and degraded ({\em middle})
snapshots. {\em Right}: Histogram representing the ratio of total
circular polarization signal in the degraded image with respect
to that in the original image. The upper $x$-axis indicates the
equivalent stray-light contamination factor.}
\label{fig:fig1}
\end{figure*}

Here we investigate in detail critical aspects of the inversion,
such as the meaning of the stray/scattered light contamination used by
Orozco Su\'arez et al.\ (2007) and the influence of noise. We also
apply our inversion strategy to {\em Hinode}/SP measurements of quiet
Sun areas at disk center. Our preliminary results suggest that most 
of the magnetic fields present in internetwork regions are weak, 
in good qualitative agreement with the picture derived from 
ground-based, near-infrared observations.

%

\section{Inversion strategy}
\label{sec:inver}

The SP aboard {\em Hinode} was designed to perform high sensitivity polarization
measurements with an unprecedented spatial resolution of 0\farcs32 and high
spectral resolution. These observations make it possible to derive the
magnetic and kinematic properties of the solar photosphere. The simplest way
to extract quantitative information from the observations is to use a
least-squared inversion technique based on ME atmospheres. In Orozco Su\'arez
et al.\ (2007) we demonstrated that the field strengths derived from ME
inversions of the \ion{Fe}{i} 630.2~nm lines observed at 0\farcs32 correspond,
on average, to the actual fields present at optical depth $\log \tau_5 \sim
-2$. In that paper we simulated {\em Hinode}/SP observations with the help of
radiative magnetoconvection simulations. The simulated spectra were inverted
in terms of simple one-component model atmospheres filling the resolution
element. The use of a {\em local} stray/scattered light profile was found to
be essential for retrieving correct magnetic field strengths.

To understand the meaning of this stray/scattered light contribution, we
resort again to the MHD simulations of V\"ogler et al.\ (2005) and Sch\"ussler
et al.\ (2003). More specifically, we consider a simulation snapshot with
unsigned average flux of 10~Mx~cm$^{-2}$.  We are certainly aware of the
limitations of current MHD calculations, but the fact is that they represent
the best option to simulate high resolution observations of quiet Sun
internetwork (IN) areas.  We have spatially degraded the polarization maps (at
each wavelength and each Stokes parameter) by telescope diffraction and
rebinned them to match the CCD pixel size of the SP (see Orozco Su\'arez et
al.\ 2007 for details). In Fig.~\ref{fig:fig1} we represent the total circular
polarization, $\int |V| \,{\rm d}\lambda/I_{\rm c}$, for the original and
degraded data. The visible pixilation corresponds to a size of
0\farcs16. There are two main effects caused by degradation: first, the
polarization signals appear to be ''blurred'' in the degraded image and,
secondly, there is a substantial loss of contrast caused by a weakening of the
polarization signals.

To evaluate more quantitatively the effects of telescope diffraction, the
right panel of Fig.~\ref{fig:fig1} displays the ratio of total circular
polarization signal in the degraded snapshot with respect to that in the
original snapshot. Only pixels whose Stokes $Q$, $U$, or $V$ amplitudes remain
larger than $4.5 \times 10^{-3} \, I_{\rm c}$ after degradation and binning to
the SP pixel size are considered here. This leaves us with 1621 pixels.  The
histogram indicates that the circular polarization is smaller in the degraded
image: 80\% of the pixels show weaker signals. The decrease in polarization
signal is not due to cancellation of opposite polarity fluxes (since mixed
polarities are not present in the snapshot at very small spatial scales), but
is truly the result of diffraction.  If one does not account for this
reduction, the inversion would systematically fail, giving too small
field strengths where the magnetic field is intrinsically weak. For this
reason it is important to use a stray/scattered light contamination factor.
Since telescope diffraction mixes light from nearby pixels, not from pixels
far away, a local stray-light profile must be considered.

\begin{figure}
\begin{center}
\FigureFile(7cm,4cm){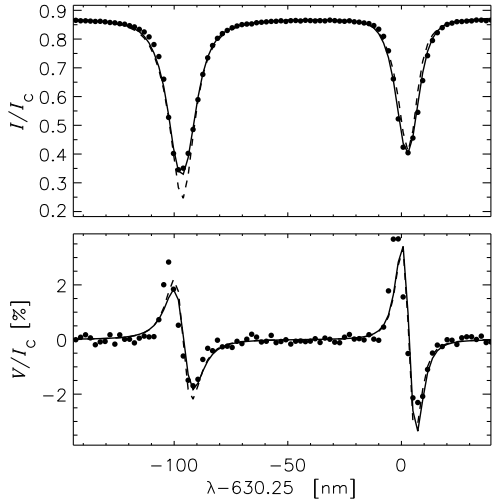} 
\end{center}
\vspace*{-1em}
\caption{Observed ({\em dots}) and best-fit Stokes $I$ and $V$ profiles
from simulated {\em Hinode}/SP observations using a global ({\em dashed})
and a local ({\em solid}) stray-light profile contamination in the
inversion.}
\label{fig:global_vs_local}
\end{figure}

Admittedly, our treatment of telescope diffraction is simplistic
because we use an {\em unpolarized} stray-light contamination, while
it is clear that diffraction also mixes the polarization signals. As a
result, pixels which show larger polarization signals after
degradation cannot be dealt with properly. However, it represents a
significant improvement over conventional treatments in which a single
global stray-light profile is employed to invert the observed region. 
This is illustrated in Fig.~\ref{fig:global_vs_local}, where
we show an example of simulated {\em Hinode}/SP observations inverted with a
global and a local stray-light profile. The best fit using a global
stray-light contamination cannot simultaneously explain the intensity
and polarization spectra because the stray-light profile has a
different shape than that needed to account for the observed Stokes
$I$ profile. The problem disappears when a local stray-light profile
is used, which also improves the determination of the intrinsic field
strength. In conclusion, even if our treatment is simple, the results
of Orozco Su\'arez et al.\ (2007) show that it does work.

\begin{figure*}
\begin{center} 
\FigureFile(16.2cm,3.6cm){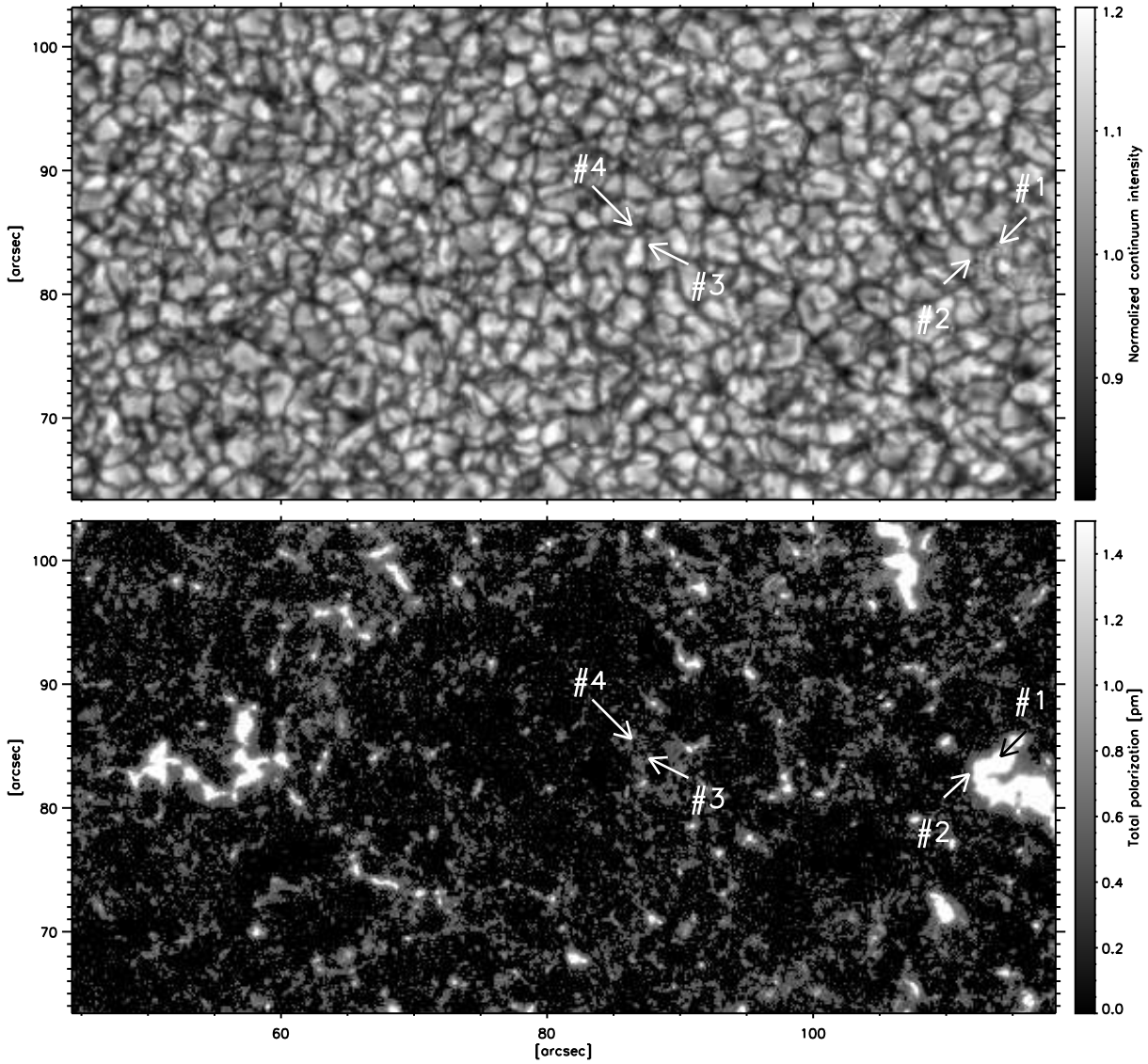} 
\end{center}
\caption{Subfield of 73.8\arcsec\/$\times$39.6\arcsec\ showing 
continuum intensities ({\em top}) and total polarization signals,
$\int (Q^2+U^2+V^2)^{1/2}\, {\rm d}\lambda/I_{\rm c}$, ({\em middle}).  The
gray scale for the total polarization saturates at 1.5~pm. The
granulation contrast is 7.44\%. Four pixel positions are indicated
with numbers. Pixels \#1 and \#2 belong to the network, while 
\#3 and \#4 are representative of IN regions.}
\label{fig:fig2}
\end{figure*}

To analyze {\em Hinode}/SP observations of quiet sun regions we propose the
approach described by Orozco Su\'arez et al.\ (2007), i.e., a least-squares
inversion technique based on ME atmospheres with simple one-component models
and a local stray-light contamination factor. As a first approximation, we
suggest to evaluate the stray-light profile individually for each pixel as the
average of the Stokes $I$ profiles observed in a box 1\arcsec\/-wide centered
on the pixel.  The inversion can be performed with only 10 free parameters:
the three components of the magnetic field (strength $B$, inclination
$\gamma$, and azimuth $\chi$), the line-of-sight velocity ($v_{\rm LOS }$),
the two parameters describing the linear dependence of the source function on
optical depth ($S_0$ and $S_1$), the line strength ($\eta_0$), the Doppler
width ($\Delta \lambda_{\rm D}$), the damping parameter ($a$), and the
stray-light factor ($\alpha$). No broadening by macroturbulence needs to be
considered. Broadening by microturbulent velocities should effectively be
accounted for by the Doppler width parameter.


\section{Observations}
\label{sec:obser}

We put the inversion strategy into practice by analyzing a quiet solar region
of 327\arcsec\/$\times$164\arcsec\/ observed with the {\em Hinode} SP at disk
center on March 10, 2007. This scan has also been studied by Lites et al.\
(2007a,b). The spectrograph slit, of width 0.16\arcsec\/, was moved across the
solar surface in steps of 0\farcs1476 to measure the four Stokes profiles of
the \ion{Fe}{i} lines at 630.2~nm. The SP samples the profiles with
2.15~pm~pixel$^{-1}$. The integration time per slit position was 4.8~s, so
completion of the map took about 3 hours. The data have been corrected for
various instrumental effects and calibrated as described by Lites et al.\
(2007c). After the reduction process, the noise levels turn out to be $1.1
\times 10^{-3} \, I_{\rm c}$ in Stokes $V$ and $1.2 \times 10^{-3} \, I_{\rm
c}$ in Stokes $Q$ and $U$, as measured in continuum wavelength regions.

Figure~\ref{fig:fig2} shows continuum intensities and total polarization
signals, $\int (Q^2+U^2+V^2)^{1/2}\, {\rm d}\lambda/I_{\rm c}$, for a small
subfield of 73.8\arcsec\/$\times$39.6\arcsec\/. The high contrast of the
granulation in the continuum intensity map testifies to the quality of the
observations. In the polarization map one can easily identify a super-granular
cell outlined by the network. The stronger polarization signals correspond to
areas where the granulation is distorted. Note that the gray scale for the
total polarization map has been clipped at 1.5~pm.  Within the cell interior,
i.e., the internetwork, we find weaker magnetic signals.  The nature of these
IN fields is a controversial topic.

\section{Inversion results}

\begin{figure*}
\FigureFile(9cm,4cm){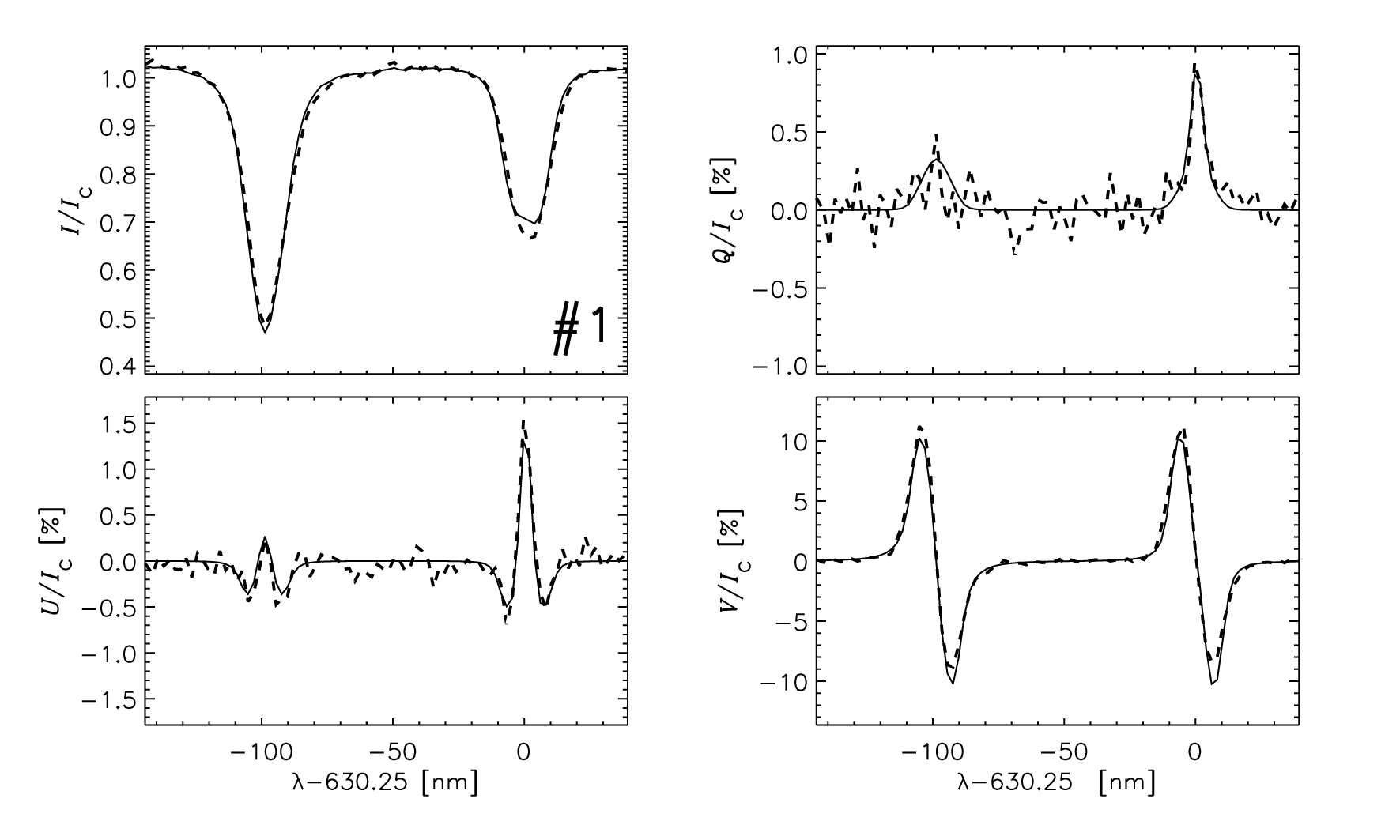} 
\FigureFile(9cm,4cm){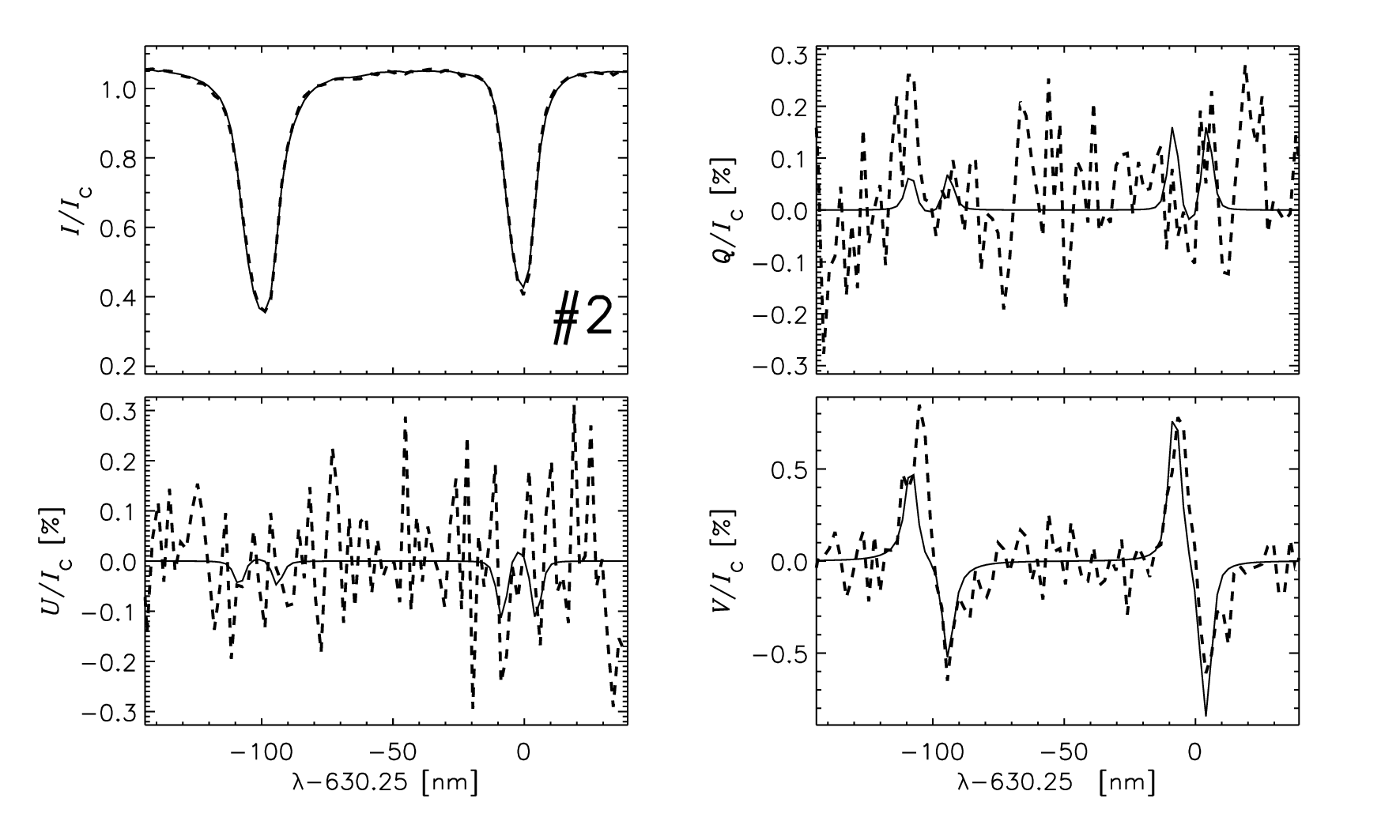} 
\caption{Observed ({\em dashed}) and best-fit ({\em solid}) Stokes profiles
emerging from network pixels \#1 ({\em left}) and \#2 ({\em right}). The field
strengths and the stray-light factors retrieved from the inversion are 1334
and 237~G, and 61 and 71\%, respectively.}
\label{fig:network}
\end{figure*}

The Stokes spectra measured by the {\em Hinode} SP have been inverted using the
MILOS\footnote{MILne-Eddington inversion of pOlarized Spectra} code (Orozco
Su\'arez \& del Toro Iniesta 2007). Figure~\ref{fig:network} displays sample
fits for individual pixels belonging to the network. Pixel \#1 ({\em left})
represents a typical network element at the center of strong flux
concentrations, whereas pixel \#2 ({\em right}) comes from the edge of 
a network patch. For both pixels the fits to Stokes $V$ are not very successful due to the
asymmetries of the profiles. Note that ME profiles are {\em symmetric} by
definition. At the edges of network patches one can see that the Stokes $V$
amplitude of \ion{Fe}{i} 630.15~nm is usually larger than that of the 630.25
nm line.  The inversion returns a field strength of 1334~G, a field
inclination of 19$^\circ$, and a field azimuth of 136$^\circ$, with a stray
light factor of 61\%, for pixel \#1, and a field strength, inclination, and
azimuth of 237~G, 69$^\circ$, and 160$^\circ$ with a stray light factor of
71\%, for pixel \#2.

\begin{figure*}
\FigureFile(9cm,4cm){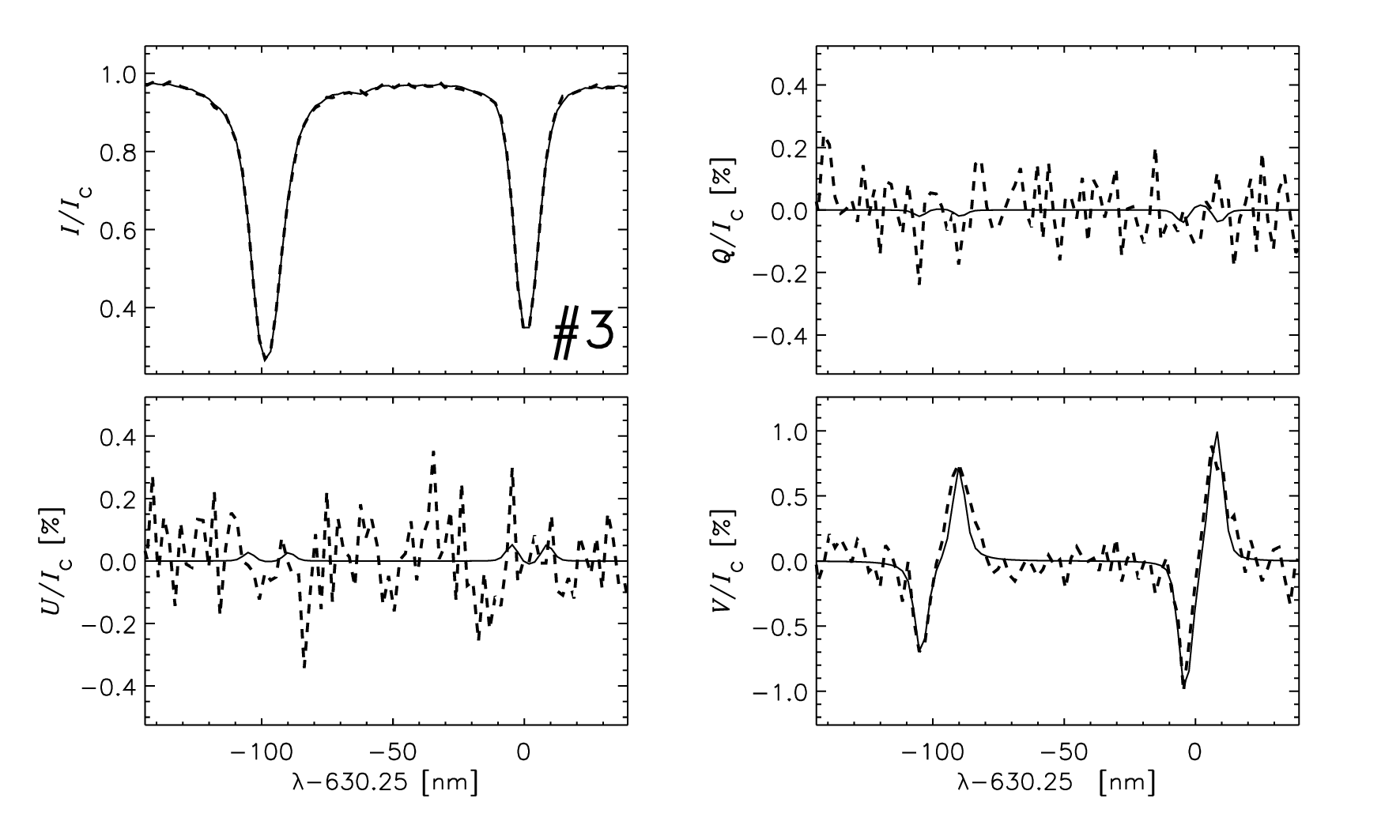} 
\FigureFile(9cm,4cm){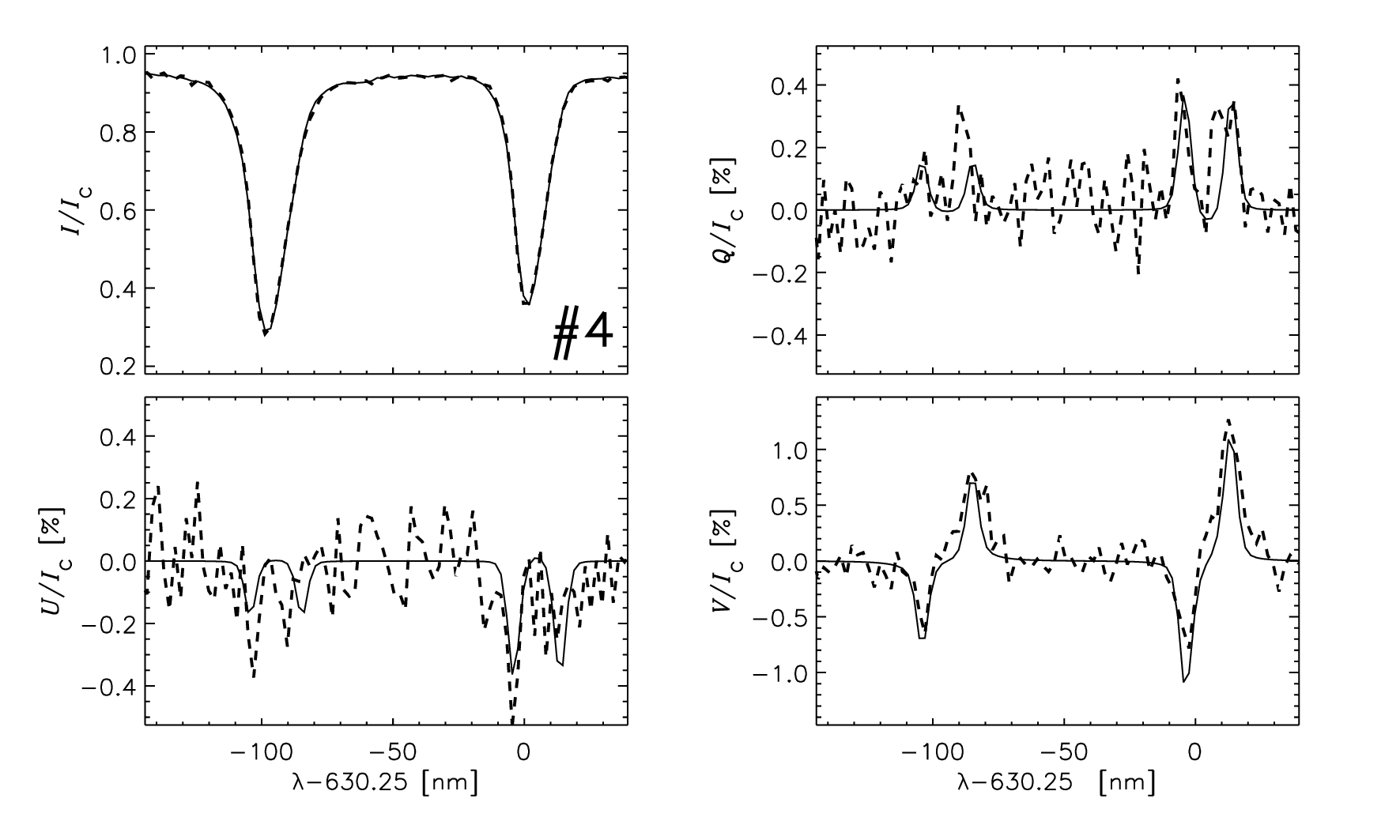} 
\caption{Observed ({\em dashed}) and best-fit ({\em solid}) Stokes profiles
emerging from the internetwork. The field strength and the stray-light factor
are 247 and 380~G for pixel \#3 ({\em left}) and 88 and 85\% for pixel \#4
({\em right}), respectively. }
\label{fig:IN}
\end{figure*}

Figure~\ref{fig:IN} displays sample profiles as well, but in this case
for two pixels belonging to the IN. The observed Stokes $V$ amplitudes
exceed $\sim$10 and $\sim$13 times the noise level, respectively. In
contrast to the profiles coming from the network, \ion{Fe}{i}
630.15~nm shows significantly smaller Stokes $V$ amplitudes than
\ion{Fe}{i} 630.25~nm, suggesting weak fields. The inversion indeed
confirms this point, retrieving field strengths, inclinations and
azimuths of 247~G, 141$^\circ$ and 248$^\circ$ for pixel \#3 ({\em
left}) and 380~G, 115$^\circ$ and 164$^\circ$ for pixel \#4 ({\em
right}).\footnote{The azimuth values are less reliable when the Stokes
$U$ and $Q$ signals approach the noise level, as in pixel \#2 (Fig.~4)
or pixel \#3 (Fig.~5).} The stray light contamination factor in the two
cases is 88\% and 85\%, respectively.

\begin{figure*}
\begin{center}
\FigureFile(16.2cm,3.6cm){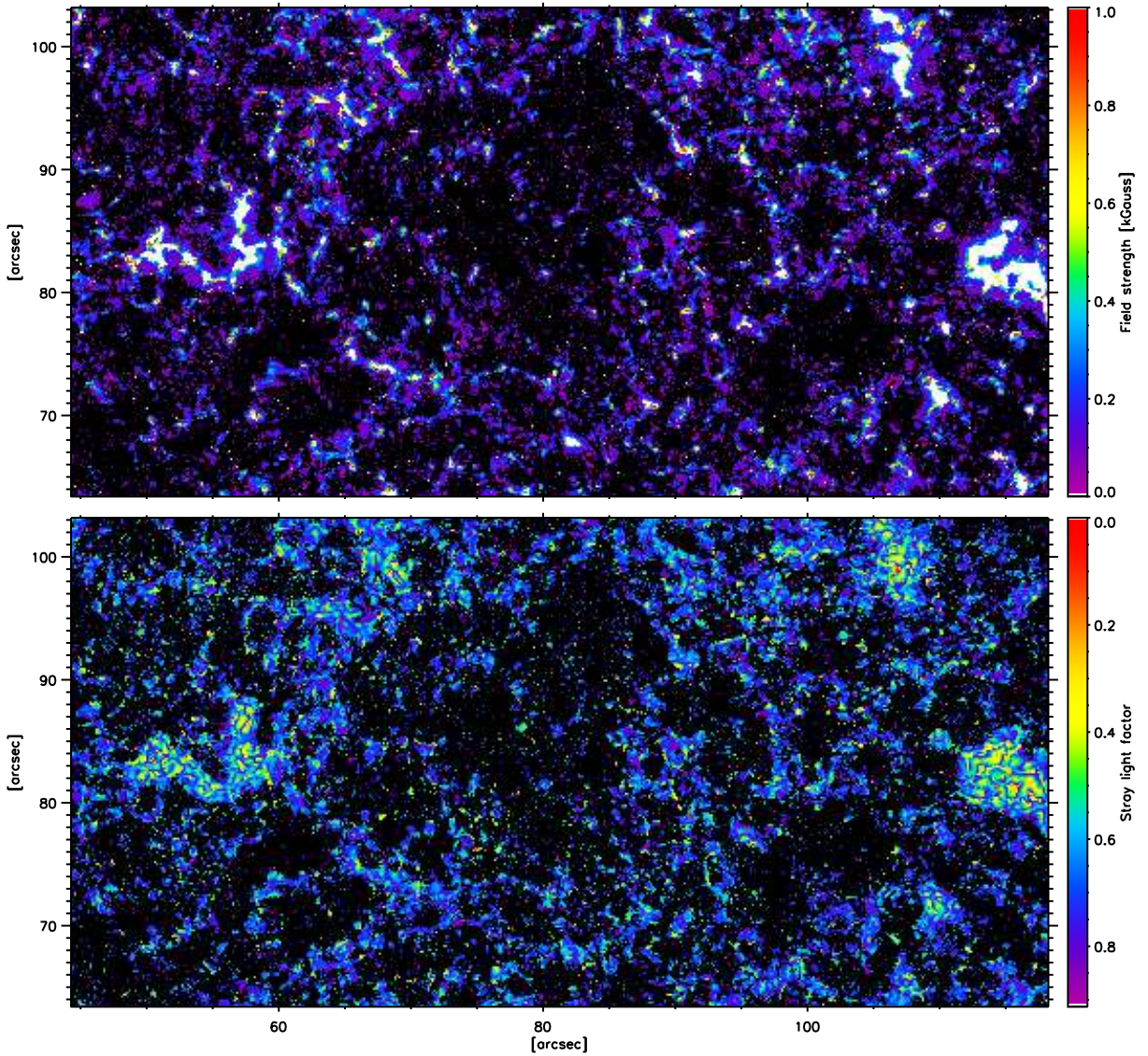}
\end{center}
\caption{Same as Fig.~\ref{fig:fig2} but for the magnetic field
strengths ({\em top}) and stray-light factors ({\em bottom})
inferred from the inversion. As in Fig.~\ref{fig:fig2}, network and
internetwork areas can easily be identified. Black areas correspond to
non-inverted pixels. The field strength color bar is clipped at
1000~G ({\em white}).}
\label{fig:fig3}
\end{figure*}

Overall, the quality of the fits is remarkably good keeping in mind
the limitations of the ME approximation and the fact that only 
one-component atmospheres are used.  

Maps of the retrieved field strengths and stray light factors are shown
in Fig.~\ref{fig:fig3}. In both maps, black regions represent pixels
whose Stokes $Q$, $U$ and $V$ signals are smaller than 4.5 times the
noise level. These pixels have not been inverted to avoid unreliable
results. As in the TCP map, two different regions can be identified:
the network, characterized by strong fields (above 1~kG), and the IN,
with much weaker fields and slightly larger stray-light factors. 

Figure~\ref{fig:fig4} displays the distribution of field strengths and
stray-light factors, for the full FOV and for IN regions, given as
probability density functions (PDFs). IN areas have been selected
manually in the interior of supergranular cells, avoiding the strong
flux concentrations of the network. The PDF of the field strength in
the full FOV peaks at about 90~G and then decreases rapidly toward
stronger fields. There are very few kG fields in this quiet Sun 
region. The PDF of the stray-light factors shows that most pixels 
require large values of stray-light contamination, with a maximum 
at around 0.8.

According to our discussion in Sect.~\ref{sec:inver}, we interpret the
stray-light contamination as a degradation of the polarization signals
due to diffraction, but it might also represent magnetic filling
factors smaller than 1. An effective filling factor can be computed as
$f = 1 - \alpha$. If one accepts this alternative interpretation, then
the fractional area of the resolution element occupied by magnetic
fields would be small, showing a peak at $f \sim 0.2$. We note,
however, that the histogram of polarization signal ratios presented in
Fig.~\ref{fig:fig1} has a maximum at around 0.5, so the stray-light
factors of $\sim 0.8$ derived from the inversion might actually
represent two different effects: telescope diffraction ($\sim$50~\%)
and a real filling factor due to still insufficient angular resolution
($\sim$30~\%).  In other words, the stray-light contaminations
required to explain the measurements of the {\em Hinode} SP may imply
an average magnetic filling factor (per pixel) of $f \sim 0.7$ in the
quiet Sun. Given that different interpretations are possible, further
work should be carried out to investigate the exact meaning of the
stray-light factors inferred from ME inversions of {\em Hinode}/SP
measurements.

%

\section{Influence of noise}

The polarization signals in the internetwork are typically smaller than those
in active regions. As a result, they are more affected by noise. This may make
the determination of vector magnetic fields less reliable. To minimize the
influence of noise we have analyzed only Stokes profiles whose polarization
signals exceed a given threshold above the noise level $\sigma$. The
inversions presented so far correspond to pixels whose $Q$, $U$ or $V$ signals
are larger than 4.5$\sigma$. This should increase the robustness of the
results because we do not include too noise profiles in the analysis. Our
experience with simulations and real observations tells us that ME inversions
provide reliable results in the quiet Sun under these conditions.

\begin{figure}
\FigureFile(8.4cm,4.2cm){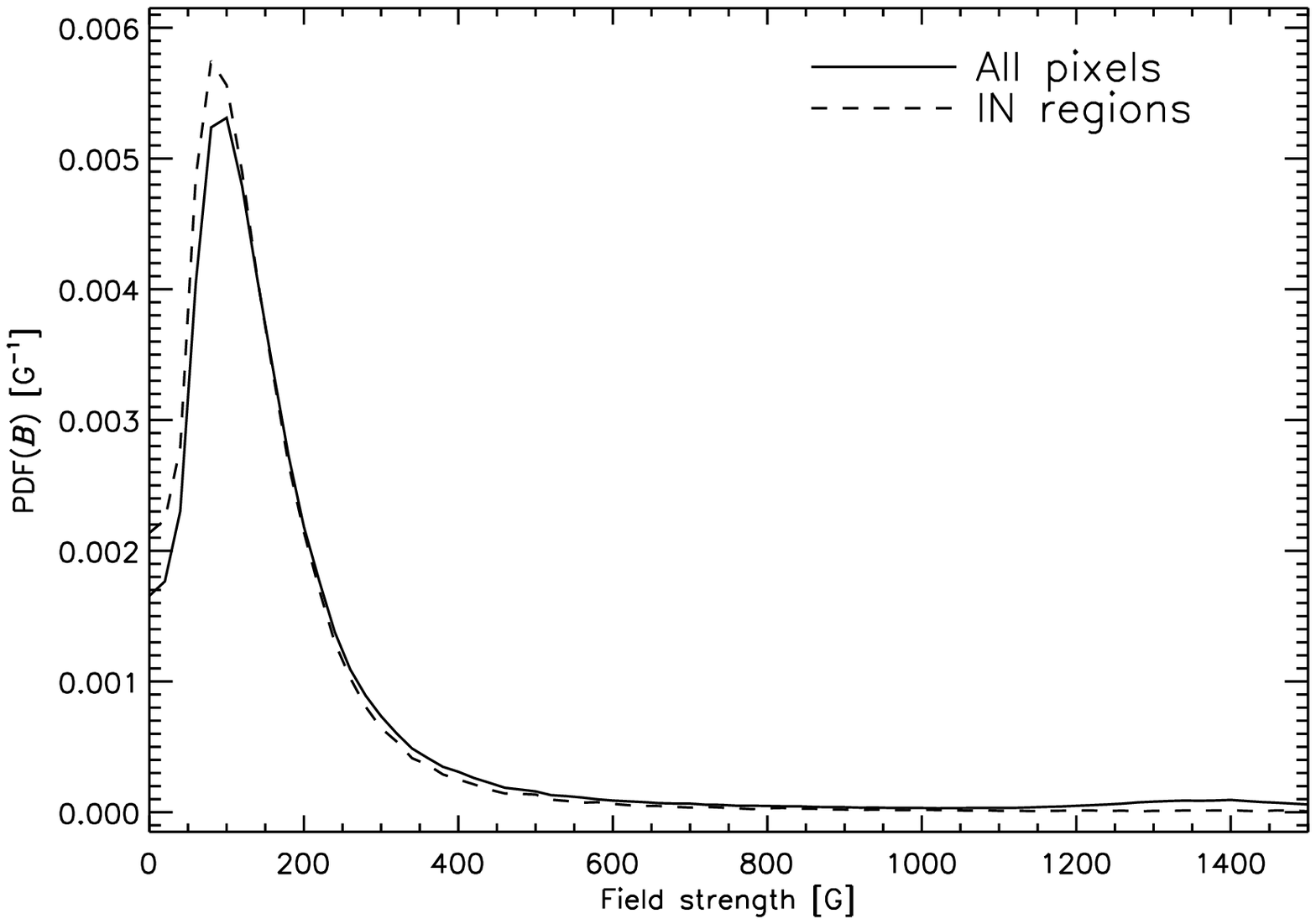} 
\FigureFile(8.4cm,4.2cm){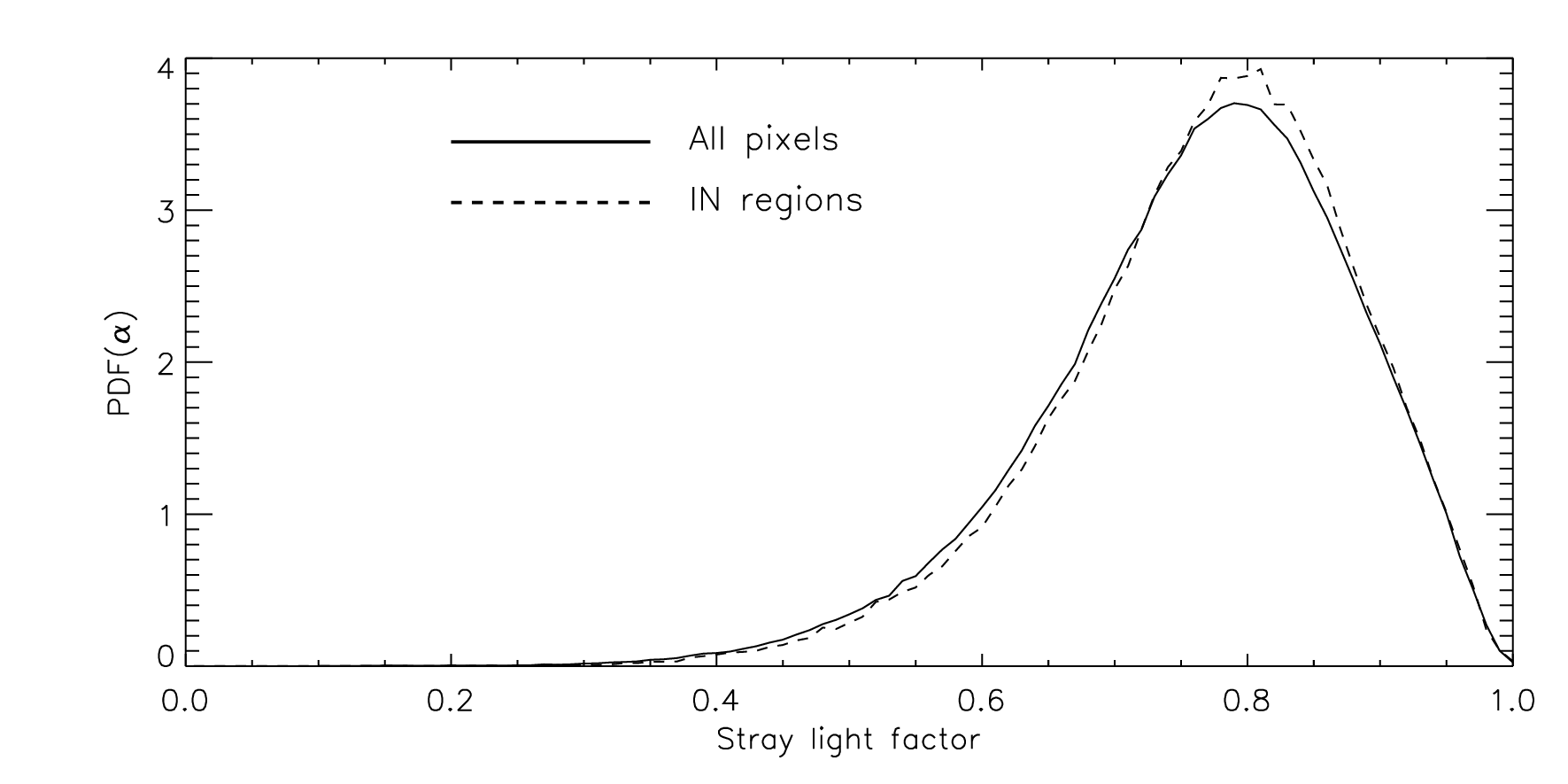} 
\caption{{\em Top}: Magnetic field strength PDF for all pixels in the FOV
and IN regions. {\em Bottom}: PDF of the stray-light factor. {\em Solid} and
{\em dashed} lines represent the full FOV and IN regions, respectively.}
\label{fig:fig4}
\end{figure}

\begin{figure}[t]
\begin{center}
\FigureFile(8.4cm,4.2cm){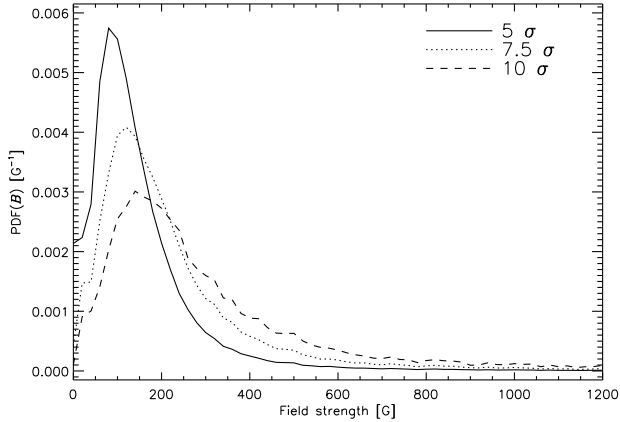}
\end{center} 
\caption{Magnetic field strength probability density
function for IN regions. The different line styles stand for different
threshold levels in the analysis. Pixels whose Stokes $Q$, $U$ or $V$
amplitudes do not exceed these levels are rejected.}
\label{fig:fig5}
\end{figure}

To evaluate the effect of noise in more detail we have calculated the field
strength PDF of IN regions for three different thresholds: 5, 7.5, and
10$\sigma$.  Figure~\ref{fig:fig5} displays the results. As the threshold
level increases, the peak of the PDF decreases in amplitude, shifts toward
stronger fields, and becomes broader. Thus, the larger the threshold, the
smaller the percentage of weak fields detected. Since weak fields are usually
associated with weak polarization signals for fully resolved magnetic
structures, this is exactly what one would expect just because the weak
polarization signals are excluded from the analysis. The important result is
that, independently of the polarization threshold used, the amount of strong
fields remains unchanged. Even for very high polarization thresholds, the
field strength PDFs are dominated by weak fields, so they cannot be the result
of noise in the profiles.

Further support to our claim that noise is not producing an artificial excess
of weak fields comes from the inversion of the deep mode observations analyzed
by Lites et al.\ (2007a,b). The deep-mode data are single-slit position
measurements with effective exposure times of 67.2 s, which lowers the rms
noise in Stokes $Q$, $U$ and $V$ down to about $3 \times 10^{-4} \, I_{\rm c}$
without compromising the spatial resolution. These observations have a S/N
about 3.7 times larger than that of the measurements considered in this
paper. A polarization threshold of 4.5 times the new noise level is equivalent
to a threshold of 1.2 times the noise level of our observations.  When we
invert the deep-mode observations with the same code and same free parameters,
we get PDFs for the field strength and stray-light factor that are nearly the
same as those presented in Fig.~\ref{fig:fig4}.

%

\section{Influence of the initial guess model}
\begin{figure}[t]
\begin{center}
\FigureFile(8.4cm,4.2cm){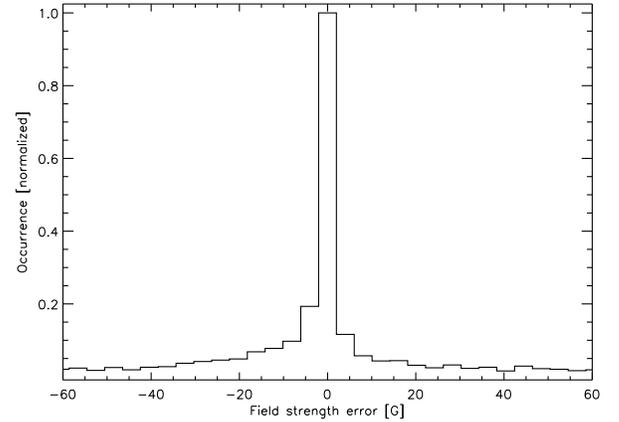} 
\end{center}
\caption{Histogram of the difference between the field strength
values inferred with two different initializations, 100 and 1500~G.}
\label{fig:fig6}
\end{figure}

Different initial guess models may lead to different results, which
has raised concerns about the uniqueness of the model atmospheres
derived from quiet Sun inversions of \ion{Fe}{i} 630.15 and 630.25~nm
(Mart\'{\i}nez Gonz\'alez et al.\ 2006a,b). The MILOS code determines
a total of ten free parameters in a maximum number of 300 iterations.
We have employed the same initial guess model for all the inversions:
$S_0=0.02$, $S_1=1$, $\eta_0=7.2$, $a=0.78$,
$\Delta\lambda_\mathrm{D}=29$~m\AA\/, $v_{\rm LOS }=0.1$~km/s,
$B=100$~G, $\gamma=30^\circ$, $\chi=30^\circ$, and $\alpha=10$\%.  How
do the results change when a strong-field rather than a weak-field
initialization is used?  To investigate this issue we have inverted a
small IN area of 32\farcs2$\times$32\farcs2 adopting different
initialization for the magnetic field strength. In particular we have
carried out four inversions with initial field strengths of 100, 500,
1000, and 1500~G. Figure~\ref{fig:fig6} shows an histogram of the
differences between the field strengths resulting from the 100 and
1500~G initializations. This plot demonstrates that the solutions do
not depend on the initial magnetic field strength.

Another indication that the results are largely independent of the initial
guess is provided by the fact that the percentage of pixels which gets
substantially better fits is small: only 3.1\% for the 500~G initialization,
4.7\% for the 1000~G initialization, and 4.3\% for the 1500~G
initialization. Here, ``substantially'' better fits mean that the final
$\chi^2$ value is at least 50\% smaller than the one obtained with 100~G.

In conclusion, even if there are unavoidable differences between the
results of different initializations, their magnitude is so small that
the field strength and stray-light distributions remain essentially
the same. This is in sharp contrast with the inversions of
ground-based measurements of the \ion{Fe}{i} 630.2~nm lines described
by Mart\'{\i}nez Gonz\'alez et al.\ (2006a). Additional
information on the robustness of inversion codes can be found in
e.g. del Toro Iniesta \& Ruiz Cobo (1996), Westendorp Plaza et
al.\ (1998), and Bellot Rubio (2006).

\section{Understanding ME inversions}
The tests presented in Sect.\ 6 demonstrate that ME inversions are capable 
of disentangling the effects of the various atmospheric parameters. In
particular, they successfully distinguish between stray-light factor and
magnetic field strength. How is this achieved in the weak field regime that
applies in most of the IN pixels? 

\begin{figure}[t]
\begin{center}
\FigureFile(8cm,4cm){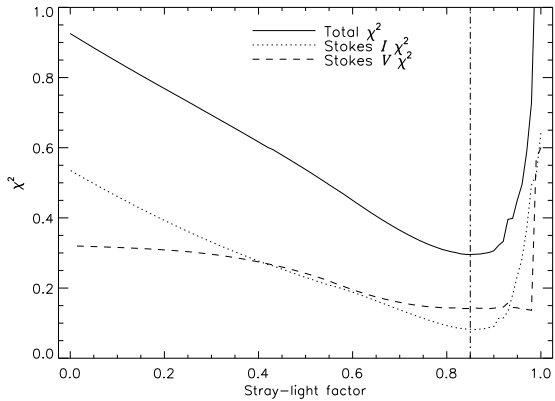} 
\end{center}
\caption{$\chi^2$ values of the best-fit
profiles resulting from the inversion of pixel \#4 (as observed
by {\em Hinode}) with fixed, erroneous stray-light factors. The
solid,dotted and dashed lines stand for the total, Stokes $I$ and $V$
$\chi^2$, respectively. The vertical line represents the ``true''
stray-light factor.}
\label{fig:fig7}
\end{figure}

To answer this question, let us assume that the ME models derived from
the inversions of Sect.\ 4 are the ``true'' solution. We have repeated
the inversion of the profiles observed by {\em Hinode} fixing the
stray-light factor to erroneous values. 101 different stray-light
factors, from $\alpha=0$ to $1$, have been considered.  The other
parameters for the initial guess model are the same as those described
in Sect.\ 6. In Fig.~\ref{fig:fig7} we represent the $\chi^2$ values
of the 101 fits against the corresponding fixed stray-light factors,
for the particular case of pixel \#4 in Fig.~\ref{fig:IN} (the results
for other pixels are equivalent). The solid line displays the total
$\chi^2$, whereas the dotted and dashed lines indicate the $\chi^2$
values for Stokes $I$ and $V$, respectively.  The vertical line
represents the ``true'' solution. Note that $\chi^2$ is dimensionless
and that its absolute value is irrelevant to the inversion code.

This plot gives a clear explanation of what is happening. When the
stray-light factor is fixed to erroneous values, the fits worsen and
the total $\chi^2$ increases. The variation of the total merit
function is large enough to be detectable by the inversion
algorithm. Remarkably, the Stokes $I$ and $V$ merit functions behave
quite differently. The Stokes $V$ $\chi^2$ values around $\alpha=0.8$
are very similar. This implies that different stray-light
contaminations produce equally good fits to Stokes $V$. In other
words: many compatible solutions, characterized by different
stray-light factors and correspondingly different field strengths,
exist for Stokes $V$.  However, the range of acceptable stray-light
contaminations is strongly limited by Stokes $I$. This is reflected in
the rapid increase of the Stokes $I$ merit function away from the
correct stray-light factor. The conclusion is the following: for the
most part, the inversion algorithm uses the information encoded in
Stokes $I$ to determine the stray-light contamination. Thus, the
often forgotten Stokes $I$ also plays an essential role in the process
of finding the absolute minimum of the total merit function.

%

\section{Conclusions}
\label{sec:con}

In this paper we have proposed a simple Milne-Eddington inversion to interpret
the high spatial resolution spectropolarimetric measurements of the quiet Sun
performed by {\em Hinode}. Using magnetoconvection simulations, we have shown that
it is important to include a stray light contamination factor in the analysis.
The stray light profile should be evaluated locally in order to account for
the effects of telescope diffraction.

The inversion strategy has been applied to a quiet Sun raster scan taken with
the {\em Hinode} SP.  We have demonstrated that noise does not significantly
affect the results of ME inversions, provided a sufficiently large
polarization threshold is used to invert the Stokes profiles. A threshold
around 4.5 times the noise level seems to yield correct inferences. In
addition, we have shown that the results do not depend on the initial magnetic
field strength of the model, because the information contained in the
Stokes profiles observed at the resolution of {\em Hinode} is sufficent to
disentangle the various model parameters.

The inferred PDFs of the magnetic field strength indicate that internetwork
regions are mainly formed by hG field concentrations with large stray light
factors. Taking into account the weakening of the polarization signals due to
telescope diffraction, these large stray light factors might also be
interpreted as magnetic filling factors of the order of 0.7. The preliminary
analysis presented here confirms the picture of weak internetwork fields
derived from ground-based measurements in the near infrared (see, e.g.,
Collados~2001).

\vspace*{2em} {\em Hinode} is a Japanese mission developed and launched by
ISAS/JAXA, with NAOJ as domestic partner and NASA and STFC (UK) as
international partners. It is operated by these agencies in co-operation with
ESA and NSC (Norway). This work has been partially funded by the Spanish
Mi\-nisterio de Educaci\'on y Ciencia through project ESP2006-13030-C06-02
(in\-clu\-ding European FEDER funds). The NAOJ Hinode Science Center is
supported by the Grant-in-Aid for Creative Scientific Research ``The Basic
Study of Space Weather Prediction'' from MEXT, Japan (Head Investigator: K.\
Shibata), generous donations from Sun Microsystems, and NAOJ internal funding.

%

\end{document}